\documentclass{article} 
\usepackage{soul}
\usepackage{amssymb}
\usepackage{mathtools, cuted}
\usepackage{commath}
\usepackage{lineno}
\usepackage{epsf,epstopdf}
\usepackage{siunitx}
\usepackage{graphicx,amssymb,amsmath}
\usepackage{txfonts}
\usepackage{mathptmx}
\begin{document}
\title{On the perturbed photogravitational restricted five-body problem: the analysis of fractal basins of convergence}
\author{Md Sanam Suraj,          Rajiv Aggarwal,        Amit Mittal,        Md Chand Asique,        Prachi Sachan}
%
%
%
%
\maketitle
\author
Corresponding author: Md Sanam Suraj

Department of Mathematics, Sri Aurobindo College, University of Delhi,  New Delhi-110017, Delhi, India

Email: mdsanamsuraj@gmail.com,  Email: mdsanamsuraj@aurobindo.du.ac.in

Abstract:

 In the framework of photogravitational version of the  restricted five-body problem, the existence and stability of the in-plane  equilibrium  points, the possible regions for motion are explored and analysed numerically, under the combined effect of small perturbations in the Coriolis and centrifugal forces.  Moreover, the multivariate version of the Newton-Raphson iterative scheme is applied in an attempt to unveil the topology of the  basins of convergence  linked with the libration points  as function of radiation parameters, and the parameters corresponding to Coriolis and centrifugal forces.


\section{Introduction}
In the recent decades, the few-body problem and particularly the circular restricted five-body problem  is one of the most fascinated as well as the challenging problem in the fields of dynamical astronomy and Celestial mechanics. The restricted problem of  five bodies describes the  dynamics of the test particle moving under the gravitational influence of the four primaries. For describing the problem in very realistic manner, the dynamics of the test particles has been studied by adding several perturbing terms in the effective potential of the circular restricted five-body problem. The study of the five-body problem with several modification is motivated by the various previous study of the restricted three or four-body problem where the same modifications are proposed and significant results are obtained (e.g., \cite{AG19}, \cite{SGA19}, \cite{AAGH17}, \cite{PE17}, \cite{AAG16}, \cite{A17},  \cite{EAKP16}, \cite{AG16}).

A particular restricted  five-body problem  is analyzed by \cite{oll88} where he explored  the dynamics of the infinitesimal body ( i.e., the fifth body) of negligible mass, in comparison to remaining four bodies, and showed the existence and stability of libration points. Further, \cite{PK07} have extended the study of same problem by including the effect of perturbation due to  the radiation of the some or even all of the primaries. The Newton-Raphson iterative scheme is applied to analyze the domain of the basins of convergence associated with the libration points  in restricted five-body problem by the Ref. \cite{ZS17}. This problem is one of the special case where $N=4$ of the  ring problem ( \cite{K99}) which defines the motion of a test particle, moving in the gravitational attraction of $N$ much bigger bodies. In which $N-1$ bodies are located in equal distances on a fictitious circle. This  ring configuration was first formulated by \cite{M90} to study the rings of Saturn.

One of the important study is the analysis of the dynamics of test particle under the effect of small perturbations in the Coriolis and centrifugal forces in various dynamical system where the primaries are also source of radiation.  Various literature is available where the effect of these forces are discussed in the restricted three as well as four-body problems.

The restricted three-body problem with the effect of small perturbation in the  Coriolis force has been studied by \cite{sze67} where he has stated that the Coriolis force is a stabilizing force.  It was observed that all the collinear libration points $L_{i}, i=1,2,3$ are unstable whereas the non-collinear libration points $L_{4,5}$ are stable when the relation $\mu_c=\mu_0+\frac{16\epsilon}{2\sqrt{69}}$ is satisfied where $\mu_0=0.03852...$, $\mu_c$ is the critical mass ratio and the change in the Coriolis force is controlled by $\epsilon$. Moreover, in the paper, only the effect of small perturbation in the Coriolis force has been considered to analyze the stability of the libration points  whereas the centrifugal force remains constant. In an extension of this study, \cite{bha78}, \cite{bha83} have discussed the stability of the libration points in the linear as well in non-linear sense by taking the effect of small perturbations in the Coriolis as well as in centrifugal forces. They further stated that  the Coriolis force is not always a stabilizing force. The collinear libration points are always unstable whereas the  non-collinear libration points are stable for all mass ratios in the range of linear stability except the three mass ratios.
Moreover, many authors have also extended their study to unveil the effect of small perturbations in the Coriolis and centrifugal forces in the frame of the restricted four-body problem  as well as in their photogravitational version, where they have discussed the  existence and stability of libration points,  regions of possible motions, and  the basins of convergence connected to the libration points of the system (e.g., \cite{agg18}, \cite{pp13},  \cite{p16}, \cite{sin15}, \cite{Sur17} ).

 Recently, the existence as well as stability of the libration points are discussed by various authors in the context of the five-body problem, i.e.,  the axisymmetric five-body problem, see Ref. \cite{gao17}, the  basins of the convergence associated with the equilibrium points, which acts as  attractors, in the axisymmetric five-body problem: the convex case,  see  Ref. \cite{sur19}, in the concave case, see  Ref. \cite{sur19b},   the five-body problem when the mass of the test particle is variable, see Ref. \cite{Sur19c} and the effect of Coriolis and centrifugal force in the axisymmetric five-body problem, see Ref. \cite{Sur19d}.

In the past few years, the Newton-Raphson basins of convergence have been studied by many authors in various dynamical system, i.e.,  the restricted three-body problem (e.g., \cite{SZK18}, \cite{Z18}, \cite{ZSMA18}),  the restricted four-body problem (e.g., \cite{Sur17, Sur17b, SMA18, SAA18}), the axisymmetric restricted five-body problem (e.g., \cite{sur19, sur19b}), restricted five-body problem  (e.g., \cite{Sur19c},  \cite{ZS17}).  The basins of convergence, linked with the libration points of the dynamical system, provide some of the most intrinsic properties of these systems.  The Newton-Raphson iterative scheme is applied to scan the set of initial conditions in an attempt to unveil the final states.  The basins of convergence is collections of all those initial conditions which converge to one of the particular attractor.

Therefore, we believe that it would be interesting to analyze the effect of small perturbations in the Coriolis and centrifugal forces on the existence as well as on the positions of the libration points in the photogravitational version of the restricted problem of five bodies.  Moreover, the effect of the radiation parameter on the basins of convergence associated with the libration points are illustrated in a systematic manner. Therefore, we believe that the present study and obtained results are novel  and this is exactly the contribution of our work.   The paper is constructed in following pattern: the Sec. \ref{Sec:2} provides the idea about of the mathematical model as well as the equations of motion of the test particle. In the following section, the positions of the libration points as function of involved parameters and their stability are discussed, whereas, the  regions of possible motion are explored in Sec. \ref{Sec:4}.  In addition, a systematic study of the  topology of the domain of  basins of convergence connected with the libration points are explored in Sec. \ref{Sec:5N}. The paper ends with Sec. \ref{Sec:6} where the discussion and conclusions regarding the main results are presented.
\begin{figure}
\centering
\resizebox{\hsize}{!}{\includegraphics{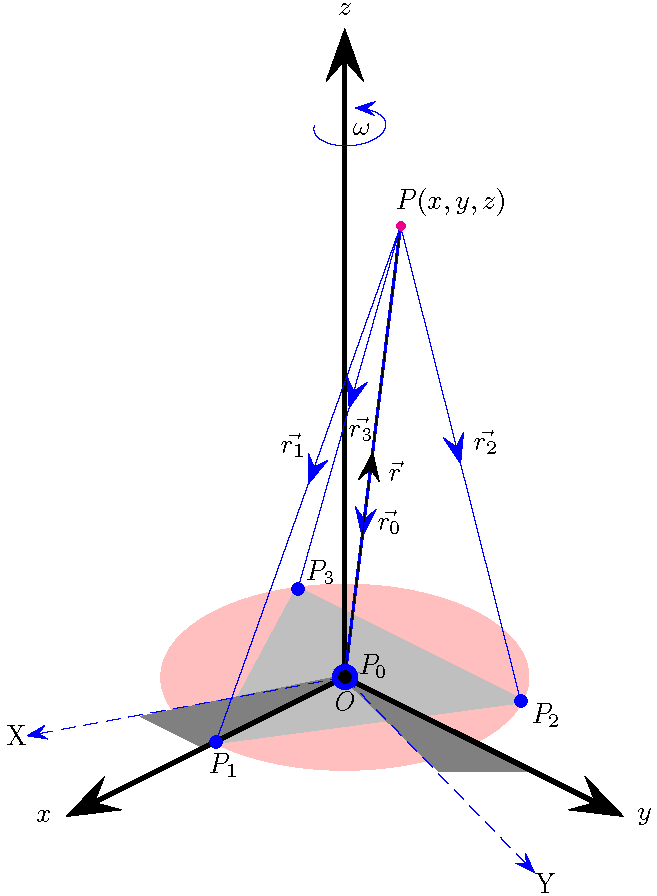}}
\caption{The planar configuration of the circular restricted five-body problem. The three bodies with equal masses $m$ are located at the vertices of an equilateral triangle, while the fourth primary, with mass $\beta m$, is located at the center of the equilateral triangle.}
\label{Fig:0}
\end{figure}
\section{Mathematical model and equations of motion}
\label{Sec:2}
The circular restricted five-body problem consists of three primaries where $P_i=1,2,3$ move in a plane under the mutual gravitational attraction in circular orbits around their common center of mass whereas the primary $P_0$ is present at the center. The fifth body, which act as an  infinitesimal test particle has a significantly smaller mass in comparison of the masses of the primaries, does not disturb the motion of the primaries.

We choose the rotating co-ordinate system in which the origin coincides  with the center of mass of the system of primaries.  The positions of the center of the primaries are:
\begin{align*}
  (x_0, y_0, z_0)&=(0, 0,0), \\
   (x_1, y_1, z_1)&=(1/\sqrt{3}, 0, 0),\\
  (x_2, y_2, z_2)&=(-x_1/2, 1/2, 0),\\
   (x_3, y_3, z_3)&=(x_2, -1/2, 0),
\end{align*}
while the dimensionless masses of the primaries are $m_0=\beta m$, $m_1=m_2=m_3=m=1$. Moreover, the three primaries $P_{1,2,3}$ with mass $m$ are placed at the vertices of an equilateral triangle with unity side, whereas the  primary $P_0$ with mass $\beta m$, is present at the center of the equilateral triangle (see Fig. \ref{Fig:0}). The line joining the centre of mass of the primaries $P_0$ and $P_1$ are taken as the $x-$axis while the line passing through origin and perpendicular to $x-$axis is taken as $y-$axis and the $z-$axis is the line passing through the origin and perpendicular to the plane of motion of the primaries.

We have further assumed that some or all the primary bodies are sources of radiation in context of the  restricted five-body problem.

We, further, apply  the transformation to scale the physical quantities, from the inertial to the synodic coordinates system. Therefore, in the rotating frame of reference, the motion of infinitesimal test particle is governed by  the following equations:
\begin{subequations}
\begin{eqnarray}
\label{Eq:1a}
\ddot{x}-2\dot{y}&=&U_x,\\
\label{Eq:1b}
\ddot{y}+2\dot{x}&=&U_y,\\
\label{Eq:1c}
\ddot{z}&=&U_z.
\end{eqnarray}
\end{subequations}
where
\begin{equation*}
U(x, y, z)=\kappa\sum_{i=0}^{3}\frac{m_iq_i}{r_i}+\frac{1}{2}\big(x^{2}+y^{2}\big).
\end{equation*}
In addition, we have  introduce the small perturbations $\vartheta$ and $\upsilon$ in the Coriolis and centrifugal  forces respectively, using the parameters $\epsilon$ and $\epsilon'$ , respectively by $\vartheta=1+\epsilon, |\epsilon|\ll1$ and $\upsilon=1+\epsilon', |\epsilon'|\ll1$. Therefore, the Eqs.  (\ref{Eq:1a}-\ref{Eq:1c}) become:
\begin{subequations}
\begin{eqnarray}
\label{Eq:2a}
\ddot{x}-2\vartheta\dot{y}&=&\Omega_x,\\
\label{Eq:2b}
\ddot{y}+2\vartheta\dot{x}&=&\Omega_y,\\
\label{Eq:2c}
\ddot{z}&=&\Omega_z,
\end{eqnarray}
\end{subequations}
where
\begin{align}\label{Eq:3}
\Omega(x, y, z)&=\kappa\sum_{i=0}^{3}\frac{m_iq_i}{r_i}+\frac{1}{2}\upsilon\big(x^{2}+y^{2}\big),\\
\kappa&=\frac{1}{3(1+\beta\sqrt{3})},\nonumber
\end{align}
and
\begin{align*}
r_{i}&=\sqrt{\tilde{x_i}^{2}+\tilde{y_i}^{2}+\tilde{z_i}^{2}}, i=0,1,2,3,\\
\tilde{x_i}&=(x-x_i), \tilde{y_i}=(y-y_i), \tilde{z_i}= (z-z_i),
\end{align*}
are the distances of the infinitesimal test particle from the respective primaries.

The effect of radiation pressure of a radiating source on a particle is expressed by the \emph{radiation factor} $q_i=1-b_i$ where $b_i$, always known as \emph{radiation coefficient} (see \cite{RZ99}), describe the ratio of the radiation force $F_{ri}$ to the gravitational force $F_{gi}$, i.e., $b_i=\frac{F_{ri}}{F_{gi}}$, thus there are four reduction factors $q_i$,  $i=0,1,...,3$.

The above mentioned problem  is a particular case of the restricted five body problem $(\beta\neq0)$ reduces to Ref. \cite{PK07} when $\vartheta=1$ and $\upsilon=1$, further reduces to Ref. \cite{oll88} when $q_i=1,  i=0,1,2,3$ while it reduces to the circular equilateral restricted four-body problem when in addition $\beta=0$.
The  Eqs.  (\ref{Eq:2a}-\ref{Eq:2c}), i.e., the equations of motion of the test particle correspond to an integral of motion, i.e.,  the Jacobi integral which reads as:
\begin{equation}\label{Eq: 4}
 J(x,y,z,\dot{x}, \dot{y}, \dot{z})=2\Omega(x,y,z)-(\dot{x}^2+\dot{y}^2+\dot{z}^2)=C,
\end{equation}
where $\dot{x}, \dot{y}$, and $\dot{z}$ represent  the velocities, while  $C$ is linked to the conserved numerical value of Jacobian constant.

\begin{figure*}
\centering
\resizebox{\hsize}{!}{\includegraphics{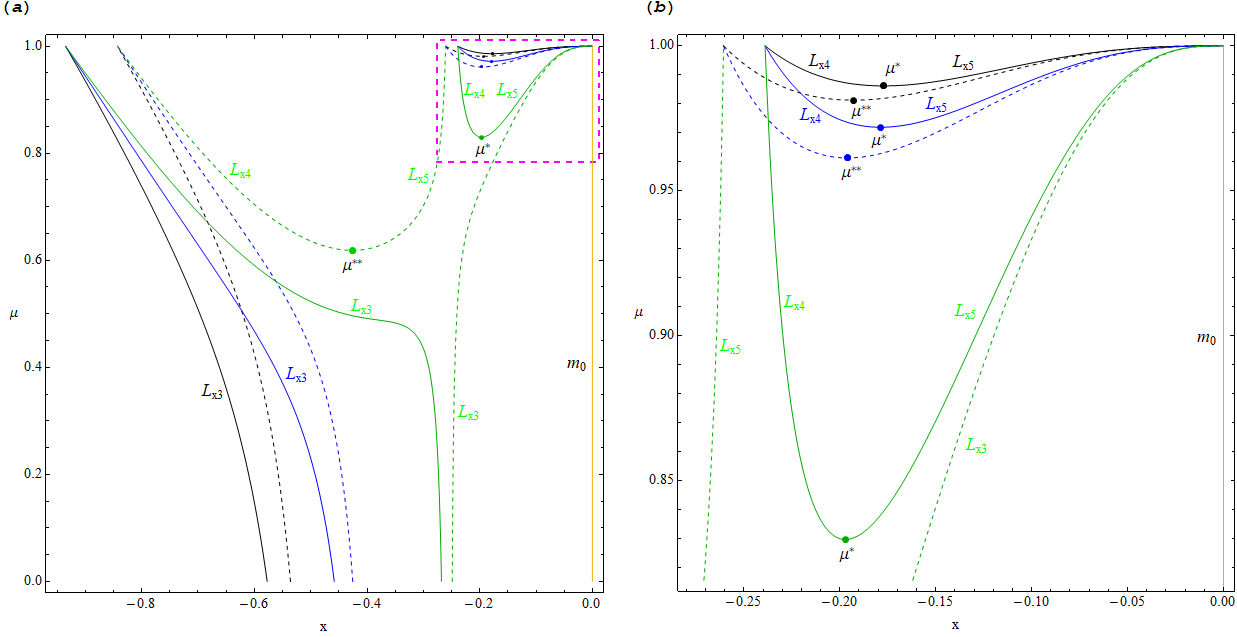}}
\caption{The  positions of the collinear libration points $L_{xi}, i=3,4,5$, on the $x-$axis as functions of $\mu$ ($q_i=1, i=1,2,3$ ).  \emph{Black} lines correspond to the gravitational case, \emph{blue} lines correspond to the case where the central primary radiate  with $q_0=0.5$ and \emph{green} lines correspond to the case where $q_0=0.1$, the solid lines  show the case when $\epsilon'=0$ while the dashed lines are representing the case when $\epsilon'=0.25$. The positions of the central primary body $m_0$ is denoted by vertical yellow lines. For blue $\mu^{**}=0.96131739,$ $\mu^{*}=0.97189778,$ and for black $\mu^{**}=0.98124858,$ $\mu^{*}=0.98617276,$ and for green $\mu^{**}=0.618699732,$ $\mu^{*}=0.82955666$. (colour figure online) }
\label{Fig:1}
\end{figure*}
To  compare, we can  define the mass parameter $ \mu= 1/(1 +\beta)$ and corresponding to the classical mass parameter of the restricted three-body problem, we have $\mu \in (0, 1]$ for $\beta \in [0, \infty)$.
\section{The libration points of the system}
We will try to illustrate how the effect of small perturbations in the Coriolis and centrifugal forces influence all the dynamical properties of the libration points.  The necessary and sufficient conditions, which must satisfy for the existence of libration points, are:
\begin{align*}
  \dot{x}= &  \dot{y}= \dot{z}= \ddot{x}= \ddot{y}= \ddot{z}=0.
\end{align*}
The linked coordinates $(x, y, z)$ of the equilibrium points can be evaluated by solving the system of following equations, numerically:
\begin{equation}\label{Eq:4}
  \Omega_x(x, y, z)=0, \Omega_y(x, y, z)=0,  \Omega_z(x, y, z)=0,
\end{equation}
where
\begin{subequations}
\begin{align}
\label{E:6a}
  \Omega_x(x,y,z) &=\frac{\partial \Omega}{\partial x}=-\kappa\sum_{i=0}^{3}m_iq_i\frac{\tilde{x_i}}{r_i^3}+\upsilon x, \\
  \label{E:6b}
  \Omega_y(x,y,z) &=\frac{\partial \Omega}{\partial y}=-\kappa\sum_{i=0}^{3}m_iq_i\frac{\tilde{y_i}}{r_i^3}+\upsilon y, \\
  \label{E:6c}
   \Omega_z(x,y,z) &=\frac{\partial \Omega}{\partial z}=-\kappa\sum_{i=0}^{3}m_iq_i\frac{\tilde{z_i}}{r_i^3}.
\end{align}
\end{subequations}
In the same vein, the second order partial derivatives of the effective potential are as follows:
\begin{subequations}
\begin{align}
\label{E:7a}
  \Omega_{xx}&=-\kappa\sum_{i=0}^{3}m_iq_i\left(\frac{1}{r_i^3}-\frac{3\tilde{x_i}^2}{r_i^5}\right)+\upsilon, \\
  \label{E:7b}
  \Omega_{yy}&=-\kappa\sum_{i=0}^{3}m_iq_i\left(\frac{1}{r_i^3}-\frac{3\tilde{y_i}^2}{r_i^5}\right)+\upsilon, \\
  \label{E:7c}
   \Omega_{zz}&=-\kappa\sum_{i=0}^{3}m_iq_i\left(\frac{1}{r_i^3}-\frac{3\tilde{z_i}^2}{r_i^5}\right),\\
   \label{E:7d}
  \Omega_{xy}&=\kappa\sum_{i=0}^{3}m_iq_i\frac{3\tilde{x_i}\tilde{y_i}}{r_i^5}=\Omega_{yx}, \\
  \label{E:7e}
   \Omega_{xz}&=\kappa\sum_{i=0}^{3}m_iq_i\frac{3\tilde{x_i}\tilde{z_i}}{r_i^5}=\Omega_{zx}, \\
    \label{E:7f}
   \Omega_{yz}&=\kappa\sum_{i=0}^{3}m_iq_i\frac{3\tilde{y_i}\tilde{z_i}}{r_i^5}=\Omega_{zy}.
\end{align}
\end{subequations}
\begin{figure*}
\centering
\resizebox{\hsize}{!}{\includegraphics{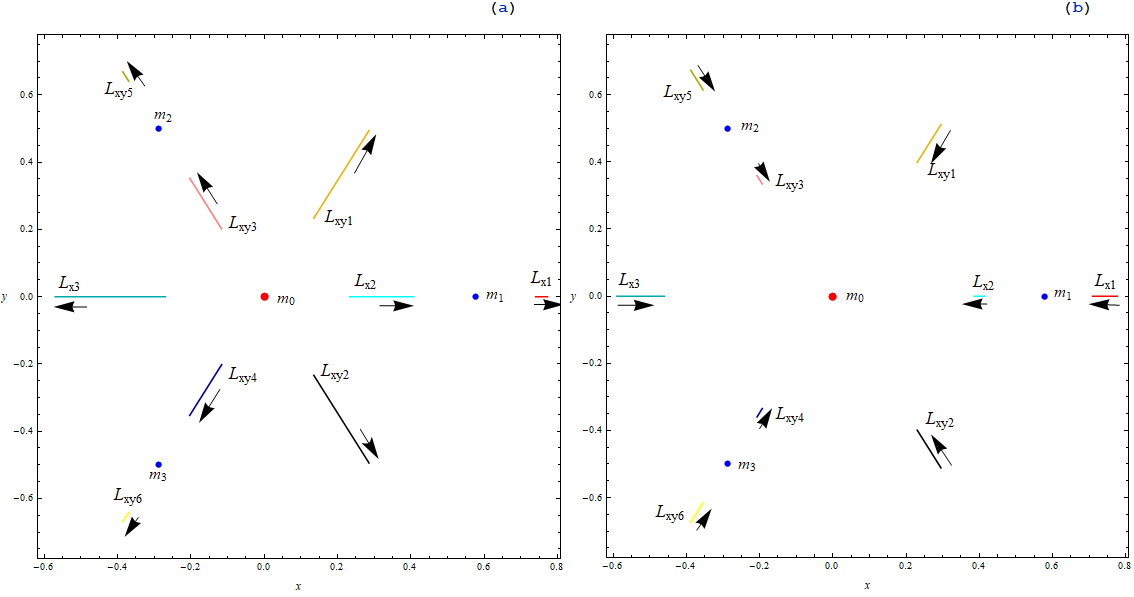}}
\caption{The parametric evolution of the positions of the libration
points, $L{xi} , i = 1, 2, 3,$ and $L_{xyi}, i=1,2, . . . ,6$ in the photogravitational circular restricted-five body problem, when $q_i=1, i=1, 2, 3,$  $\mu=0.09$; and  (a) $q_0 \in (0, 1]$, $\epsilon' =0$ ; (b) $\epsilon' \in [0,1)$, $q_0=1$. The black arrows indicate the movement direction of the libration points as the value of the  corresponding parameters  increase. The blue and red dots pinpoint the fixed centers of the primaries $P_i, i=1,2,3,$ and the radiating primary $P_0$, respectively.}
\label{Fig:02}
\end{figure*}
\begin{figure*}
\centering
\resizebox{\hsize}{!}{\includegraphics{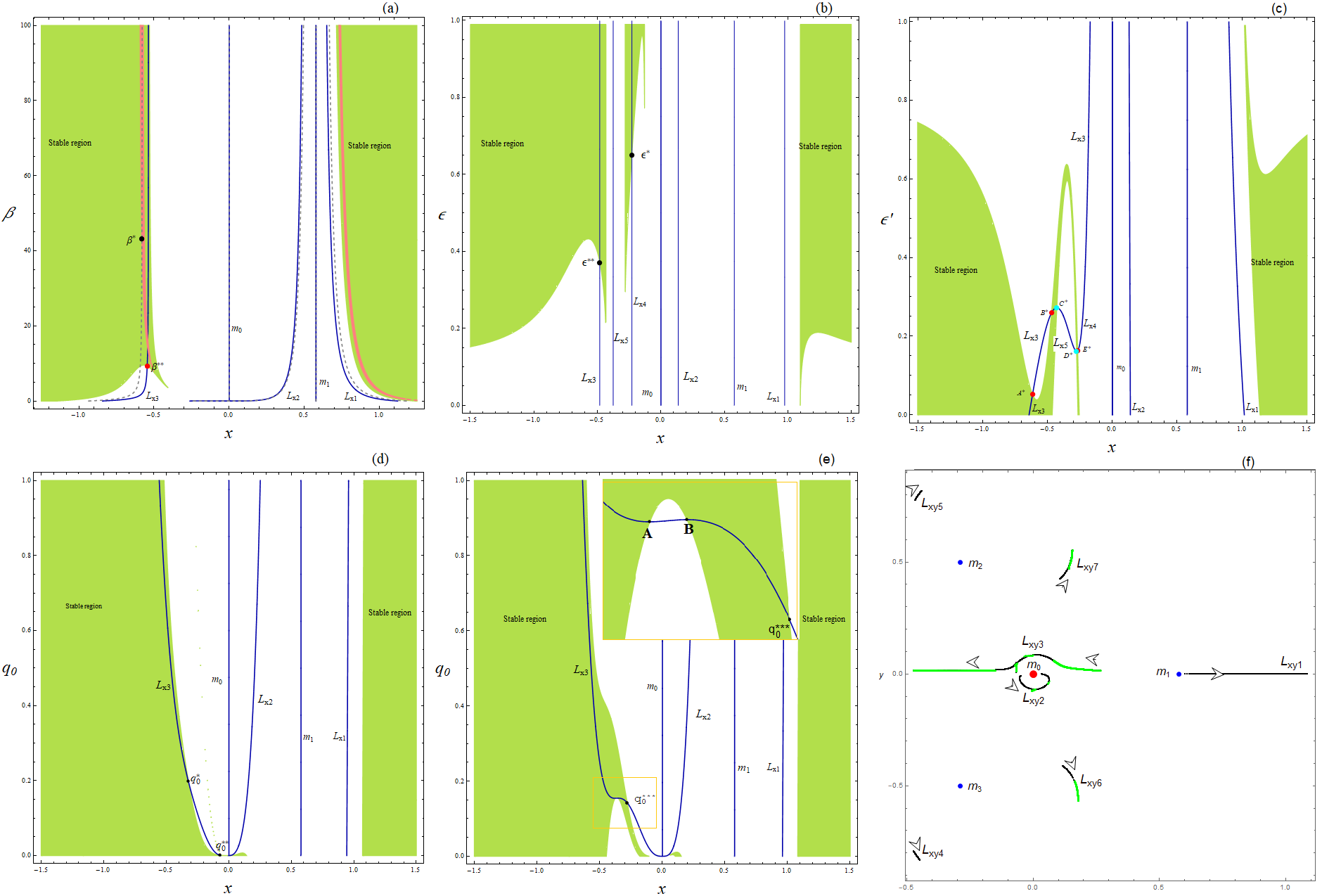}}
\caption{The stability diagram for the collinear libration points  for $q_i=1, i=0,1,2,3$, and (a); the \emph{black} and \emph{red} colour dots show the critical value of $\beta$ i.e., $(L_{x3}, \beta^*)=(-0.5803558702, 43.1810594751)$; and $(L_{x3}, \beta^{**})=(-0.5451484653, 9.3205312844)$;  for the classical case and the case when $\epsilon=\epsilon'=0.25$ respectively in restricted five-body problem. The \emph{gray} dashed lines show the positions of collinear libration points in the classical case while the solid blue lines show the positions of the libration points in the case when the effect of small perturbations in the Coriolis and centrifugal forces have been considered. (b); the stability diagram for $\mu=0.628699732$, $q_0=0.1$, $\epsilon'=0.25$ and the black dots show the critical value of $\epsilon$ where $(L_{x3}, \epsilon^{**})=(-0.481457, 0.370814)$, $(L_{x4}, \epsilon^{*})=(-0.227775, 0.65071)$;
(c); $q_i=1, i=1,2,3,$; $q_0 = 0.1$; $\epsilon= 0.35$; $\mu = 0.628699732$; $A^*=0.05138987$, $B^*=0.25961479$, $C^*=0.27220018$, $D^*=0.161740147$, $E^*=0.16317148$,
(d) $q_1=0.95$, $q_2=0.65$, $q_3=0.65$, $\epsilon=0.55$, $\epsilon'=0.35$, and $\mu=0.628699732$; $q_0^*=0.19913869$, $q_0^{**}=0.00089653$,  (e) $q_i=1,$ $i=1,2,3,$ and $\epsilon=0.55$, $\epsilon'=0.35$, $\mu=0.628699732$;  the critical values of $q_0$ is denoted by $q_0^{***}=0.14242323$; $A=0.15431122$, and $B=0.15457255$ (f); $\mu=0.98124858$, $\epsilon'=0.25$, $q_0=0.15$,  $q_2=0.35$,  $q_3=0.4$,  and the radiation parameter $q_1\in(0,1]$.}
\label{Fig:S1}
\end{figure*}
\begin{figure}
\centering
\resizebox{\hsize}{!}{\includegraphics{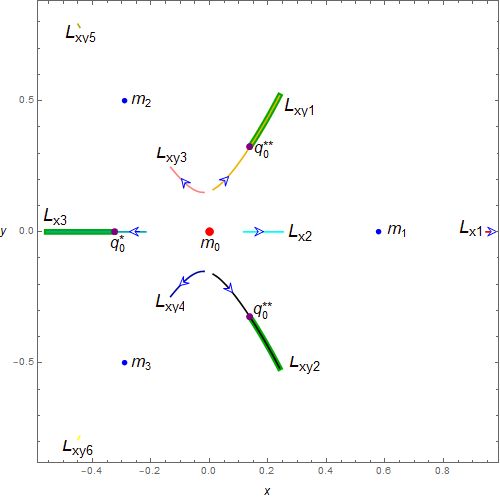}}
\caption{The stability diagram for the  libration points for $q_1 = 0.95$; $q_2 = 0.65$;  $q_3 = 0.65$; $\varepsilon= 0.55$; $\varepsilon' = 0.35$; $\mu= 0.628699732$; and  $q_0\in[0.08,1]$ the critical value of $q_0$ for $L_{x3}$ is  $q^{*}_0 = 0.19913869$ and for $L_{xy1, 2}$ is $q^{**}_0 = 0.28494$. The \emph{blue} arrows show the movement of the libration points while the \emph{purple} dots show the critical value of $q_0$. The \emph{green} thick lines show the stable libration points. (colour figure online)}
\label{Fig:S2}
\end{figure}
More precisely, when we restrict this study only  to  the coplanar libration points, i.e.,  the points which are located on the configuration $(x, y)$ plane with $z=0$.

 In addition for simplicity, we consider only the central primary is radiating, i.e., $q_i=1, i=1,2,3$. Therefore, the intersections of the nonlinear equations $\Omega_x(x,y,0)=0$ and $\Omega_y(x,y,0)=0$ define the locations of the coplanar libration points.  In Ref. \cite{PK07}, where the effect of the Coriolis and centrifugal forces has been neglected, it is shown that the critical value of the mass parameter $\mu$ is function of $q_0$, i.e., the radiation factor when the central primary $P_0$ is source of radiation. When $q_0=1$,  then $\mu^*=0.98617276$, and when $q_0=0.5$,  then $\mu^*=0.97189778$, and when $q_0=0.1$,  then $\mu^*=0.82955666$, therefore it is concluded that when the radiation factor $q_0$ increases, the value of $\mu^*$ increases. Where the $\mu^*$ is the critical value where the number of collinear libration points changes. The problem admits three collinear libration points for $\mu\in(0, \mu^*)$ on the $x-$axis in which $L_{x1}$ lies out side the primary $m_1$ whereas the libration point $L_{x2}$ lies between $m_0$ and $m_1$ and the libration point $L_{x3}$ lies on negative $x-$axis. For $\mu\in[\mu^*, 1)$ the problem admits two more collinear libration points $L_{x4,5}$ when $\epsilon'=0$, i.e., the effect of centrifugal force is neglected and therefore there exists fifteen libration points in total.
\begin{figure*}
\centering
\resizebox{\hsize}{!}{\includegraphics{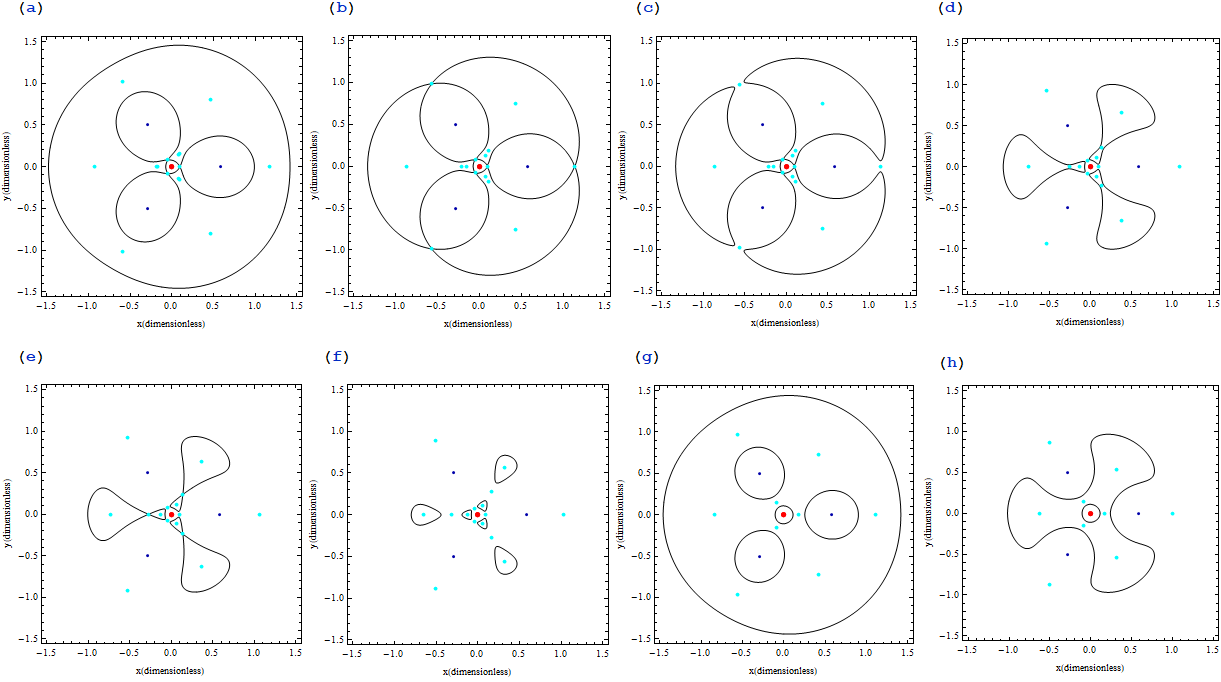}}
\caption{The zero velocity curves for $\mu=0.98627276, q_i=1, i=0,1,2,3$, and $C=3.519767801$; (a) $\epsilon'=0$; (b) $\epsilon'=0.15$; (c) $\epsilon'=0.153$; (d) $\epsilon'=0.52$; (e) $\epsilon'=0.59$; (f)$\epsilon'=0.9$, and with $\mu=0.89$, (g)$\epsilon'=0.1$; (h) $\epsilon'=0.8$. The \emph{blue} colour dots show the positions of the three primary and \emph{red} dot shows the radiating primary while the \emph{teal} colour dots show the positions of the libration points.}
\label{Fig:Z1}
\end{figure*}
\begin{figure}
\centering
\resizebox{\hsize}{!}{\includegraphics{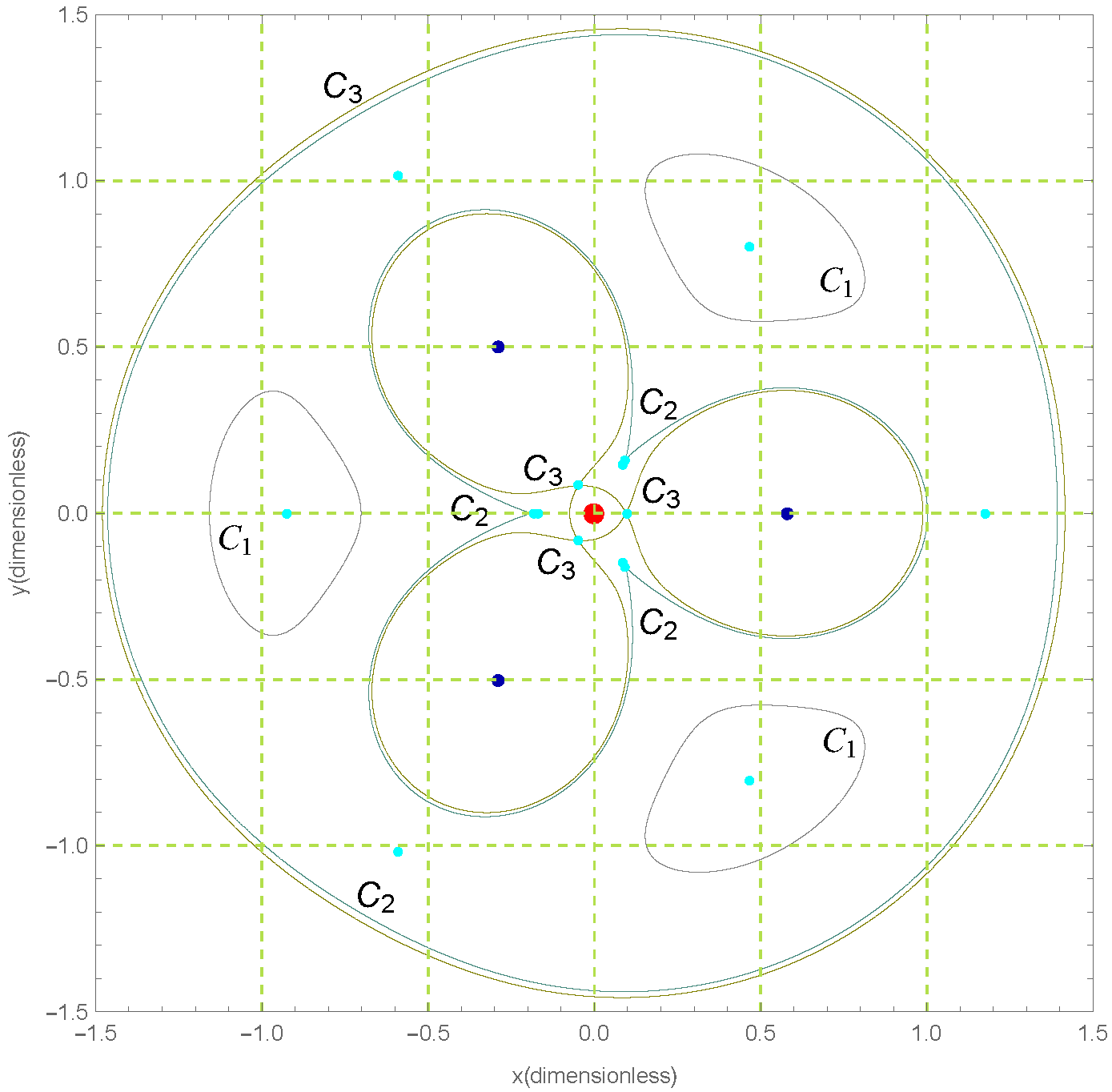}}
\caption{The zero velocity curves for $\mu=0.98627276, q_i=1, i=0,1,2,3$; $\epsilon'=0$; and $C_1=3.021930316$; $C_2=3.48676523$; $C_3=3.519767801$. The \emph{blue} colour dots show the positions of the three primary and \emph{red} dot shows the radiating primary while the \emph{teal} colour dots show the positions of the libration points.}
\label{Fig:Z2}
\end{figure}

In Fig. \ref{Fig:1}, we have depicted how the positions of the collinear libration points $L_{x3,4,5}$ change when the radiation parameter as well as the parameter $\epsilon'$ which controlled the  centrifugal force change. It is observed that as the value of $q_0$ increases, the position of  libration point $L_3$ move far from the origin whereas the position of this  libration point moves towards the origin when $\epsilon'$ increases. When we have taken the effect of the  centrifugal  force into the consideration, it can be easily observed that for $\mu^{**} < \mu^*$ in all the considered three cases which means that the interval $\mu^\in(0, \mu^{**}] $  there exist only three collinear libration points whereas for $\mu^\in(\mu^{**}, 1)$,  two more libration points occur. Therefore, the interval shrinks where the nine libration points exist and consequently the interval increases in which the fifteen libration points exist when the effect of the  centrifugal  force is taken into consideration.

It is observed that circular restricted five body problem has an  axis of symmetry, i.e., x-axis. Indeed, if the $x-$axis is rotated successively through an  angle of $2\pi/3$, two additional lines of symmetry $y=\pm\sqrt{3}$ are obtained. We have observed that  two more symmetry lines of the problem exist where three or five libration points exist  on each of them with respective positions as the ones on the $x-$axis. Therefore, we have a total of nine or fifteen libration points  of the problem when $q_i=1,  i=1,2,3$.

In Fig. \ref{Fig:02}, the movement of the positions of the libration of points are presented, where Fig. \ref{Fig:02}a, is illustrated for varying value of radiation parameter $q_0$, and Fig. \ref{Fig:02}b, is illustrated for varying value of parameter $\epsilon'$, i.e., the effect of perturbation in the coriolis force parameter. It is further observed that as the radiation parameter $q_0$ increases all the nine libration points $L_{xi}, i=1,2,3$ and $L_{xyi}, i=1,2,...,6$ move far from the central primary whereas all these libration points move towards the central primary when the parameter $\epsilon'$ increases.

In the case when the radiation pressure due to other primaries are taken into consideration i.e., $q_i\neq1,  i=1,2,3$,  there exist no collinear libration points when $q_2\neq q_3$. It is observed that when $q_2 \neq q_3$ there exist only five or seven non collinear libration points depending on the combination of the parameters $\epsilon', q_i, i=0,1,2,3$.

From Eq. \ref{Eq:3},  it is observed that the first order  partial derivatives of $\Omega$ are free from the parameter $\epsilon$ controlling the Coriolis force. Therefore, the positions of the coplanar libration points remains unaffected by this perturbation parameter. However, this parameter effect significantly the stability of these libration points.

The study of the stability of  libration points in any dynamical system plays a crucial role to unveil the various properties of that system. In an attempt to study the linear stability of the coplanar libration points $(x_0,  y_0)$, a displacement is given to the positions of  infinitesimal mass as $(x_0+\phi, y_0+\varphi)$ where $\phi$, and $\varphi$ denote the perturbations along the $ox$ and $oy$ axes respectively.

Therefore, expanding the equations of motion $(\ref{Eq:2a}$ and  $\ref{Eq:2b})$ into first-order terms, with respect to $\phi$ and $\varphi$, we have
\begin{subequations}
\begin{eqnarray}
\label{Eq:8a}
  \ddot{\phi} -2\vartheta\dot{\varphi}&=& \phi{\Omega_{xx}}_0+ \varphi{\Omega_{xy}}_0,\\
  \label{Eq:8b}
  \ddot{\varphi}+2\vartheta\dot{\phi} &=&   \phi{\Omega_{yx}}_0 + \varphi{\Omega_{yy}}_0,
\end{eqnarray}
\end{subequations}
where ${{\Omega_{xx}}_0, \Omega_{yy}}_0$, and ${\Omega_{xy}}_0$ are the second order derivatives of $\Omega$, with respect to $x$ and $y$, evaluated at the libration points.

Substituting the solutions of the variational equations, i.e., $\phi=\xi_1e^{\lambda t}$ and $\varphi=\xi_2 e^{\lambda t}$ in Eqs. (\ref{Eq:8a}, \ref{Eq:8b}),  we obtain:
\begin{subequations}
\begin{eqnarray}
\label{Eq:9a}
(\lambda^2-{\Omega_{xx}}_0)\xi_1-(2\vartheta\lambda+{\Omega_{xy}}_0)\xi_2&=&0,\\
 \label{Eq:9b}
 (\lambda^2-{\Omega_{yy}}_0)\xi_2+(2\vartheta\lambda-{\Omega_{yx}}_0)\xi_1&=&0.
 \end{eqnarray}
\end{subequations}
The Eqs. (\ref{Eq:9a}, \ref{Eq:9b}) have nontrivial solution if the determinant of the coefficients matrix of the system vanishes. Therefore, the characteristic equation associated with the system of linear equations  (\ref{Eq:9a}, \ref{Eq:9b}) is quadratic in $\Xi=\lambda^2$ and given as follows:
\begin{equation}\label{Eq:10}
\wp_1\Xi^2+\wp_2\Xi+\wp_3=0,
\end{equation}
where
\begin{subequations}
\begin{eqnarray}
\label{E:11a}
  \wp_1 &=& 1, \\
  \label{E:11b}
  \wp_2 &=& 4\vartheta^2- {\Omega_{xx}}_0-{\Omega_{yy}}_0,\\
  \label{E:11c}
  \wp_3 &=& {\Omega_{xx}}_0 {\Omega_{yy}}_0-{\Omega_{xy}}_0{\Omega_{yx}}_0.
\end{eqnarray}
\end{subequations}
The associated libration points are said to be stable if all the four roots of the characteristic equation are pure imaginary. This happens only when the conditions, illustrated below, satisfy simultaneously:
\begin{equation}\label{Eq:12}
  \wp_2>0,\quad  \wp_3>0,\quad \wp_2^2-4\wp_1\wp_3>0.
\end{equation}
Indeed, if $\lambda^2=\Xi$ then the characteristic equation has two real negative roots which consequently give four pure imaginary roots in $\lambda$.

In Fig. \ref{Fig:S1}, the stability diagrams of the collinear libration points are illustrated. In Fig. \ref{Fig:S1}a, the stability diagram is presented for the case when none of the primary is radiating. The \emph{pink} colour region shows the stability region for the classical case of restricted five-body problem. It can be noticed that the only most negative libration point on the $x-$axis i.e., $L_{x3}$ is stable for $\beta^*=43.1810594751$, where $\beta^*$ is the critical value of $\beta$ such that the libration point $L_{x3}$ is stable for $\beta\geq \beta^*$. The green colour region shows the stability region for the case when the effect of small perturbations in the Coriolis and centrifugal forces have been taken into account. It is observed that the stable regions increase in comparison of the classical case (see \cite{oll88}) and also the $\beta^{**}=9.3205312844<\beta^*$ i.e., the critical value of $\beta$ also decreases significantly in comparison of the $\beta^*$. However, for this set of values only the libration point $L_{x3}$ is stable. In Fig. \ref{Fig:S1}b, we have illustrated the stability region when only the  central primary is source of radiation. It is observed that as the effect due to small perturbation in the Coriolis force  increases the libration points $L_{x3,4}$ become stable.  The critical value of $\epsilon$ are denoted by $\epsilon^*$ and $\epsilon^{**}$ such that the libration point $L_{x3}$ is stable for $\epsilon \geq \epsilon^{**}$ while $L_{x4}$ is stable for $\epsilon \geq \epsilon^*$. Therefore, it can be concluded that the libration point $L_{x4}$  is stable for slightly higher value of $\epsilon$. In Fig. \ref{Fig:S1}c, the stability regions for the collinear libration points are discussed for varying value of the parameter $\epsilon'$ and fixed value of the other parameters. It is observed that the collinear libration point $L_{x3}$ is stable for the two different intervals.

In Fig. \ref{Fig:S1}d, the stability region is depicted for varying value of the radiation parameter $q_0$ where the other primaries are also source of radiation. It is observed that the libration point $L_{x3}$ is stable in two intervals,  i.e., $q_0\in (0, q_0^{**}]$ and $q_0\in[ q_0^{*}, 1]$.

In Fig. \ref{Fig:S1}e, the stability region is illustrated only when the central primary is radiating. It is observed that the collinear libration point $L_{x3}$ is stable for the two different intervals, i.e., $q_0\in[q_0^{***}, B]$ and $q_0\in[A,1]$.

 The movement of the positions of the libration points has been illustrated in Fig. \ref{Fig:S1}f, where the radiation parameter of each primary is different, i.e., $q_0=0.15$,  $q_2=0.35$,  $q_3=0.4$,  and the radiation parameter $q_1\in(0,1]$.  It can be observed that there exist only non-collinear libration points and the number of libration points varies from five to seven as the value of the radiation parameter $q_1$ due to the primary $P_1$ increases. The stability analysis shows that the libration points $L_{xy2,3,6,7}$ are stable in some interval of radiation parameter $q_1$. The stable libration points are shown in green colour.

For $q_i=1, i=1,2,3$; $q_0= 0.1$; $\epsilon'= 0.25$;  and $\mu= 0.628699732$ there exist fifteen libration points in which five are collinear i.e., $L_{xi}, i=1,2,...,5$ and ten are non-collinear i.e., $L_{xyi}$, $i=1,2,...10$. When we have analyzed the effect of the parameter $\epsilon$ on the stability of the these libration points, it is unveiled that the libration points $L_{x3}$ and $L_{xy1,2}$ are stable for $\epsilon\in[\epsilon^{**}=0.370814,1]$  while $L_{x4}$ and $L_{xy7,8}$ are stable for $\epsilon\in[\epsilon^{*}=0.65071,1]$.

For $q_i=1, i=1,2,3$; $q_0= 0.1$; $\epsilon= 0.35$; $\mu= 0.628699732$ and varying value of $\epsilon'$, it is observed that in the interval $\epsilon' \in(0, D^*] $ and $(C^*, 1]$ there exist nine libration points whereas for $\epsilon' \in(D^*, C^*] $ there exist fifteen libration points. The numerical investigations for  the stability of these libration points unveil that the libration points $L_{x3}, L_{xy1, 2}$ are stable in  $\epsilon' \in(0, A^*] $, $L_{x5}, L_{xy1, 2}$ are stable in $\epsilon' \in(B^*, C^*] $ and $L_{x5}, L_{xy7, 8}$ are stable in  $\epsilon' \in(D^*, E^*] $.

In  figure \ref{Fig:S2}, the parametric evolution of the  movement of position of libration points are illustrated when the radiation parameters for the primaries $m_{2,3}$ are same and for varying value of radiation parameter $q_0\in(0.08,1]$. It can be observed that the positions as well as the movement of the non-collinear libration points are symmetrical about the $x-$axis. Also, all the libration points move far from the central primary as the value of the parameter $q_0$ increases. The thick \emph{green} lines show the stable region for the libration points $L_{x3}$ and $L_{xy1,2}$.  The non-collinear libration points  $L_{xy1,2}$ are stable for $q_0\in(0.28494, 1]$ with the same set of values for the other parameters while the co-linear libration point $L_{x3}$ is stable for $q_0\in(q_0^*, 1)$.
\section{The regions of possible motion}
\label{Sec:4}
\begin{figure*}
\centering
\resizebox{\hsize}{!}{\includegraphics{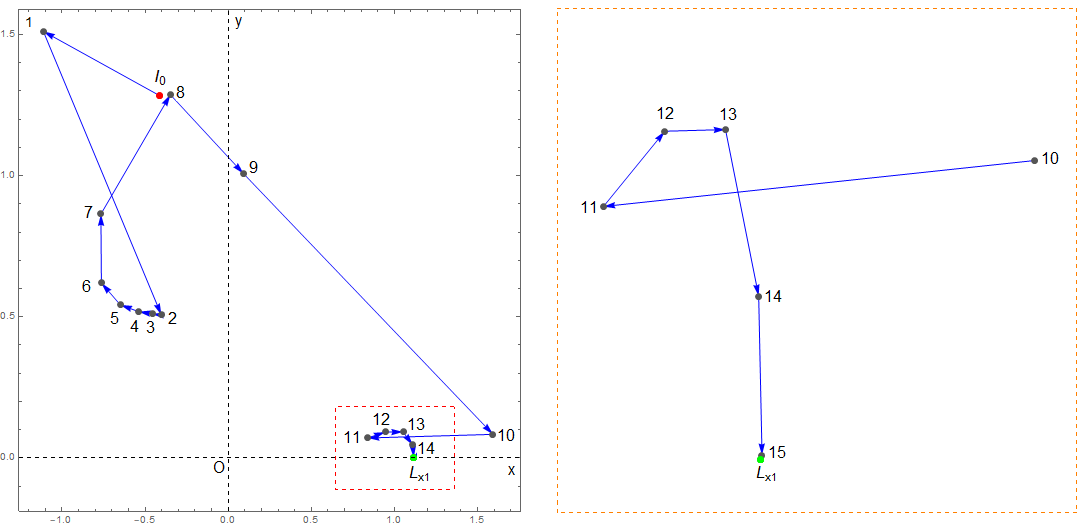}}
\caption{A characteristic example of consecutive steps that are followed by the Newton-Raphson iterative scheme and the associated crooked path-line that leads to a libration point $L_{x1}$. The starting point $I_0(x_0, y_0)=(-0.4146, 1.284)$ is depicted by \emph{red} dot while the libration point is pin pointed by green dot, to which the method converges. The Newton-Raphson method converges after 16 iterations to $L_{x1}$ with accuracy of six digits.}
\label{Fig:CP1}
\end{figure*}
\begin{figure*}
\centering
\resizebox{\hsize}{!}{\includegraphics{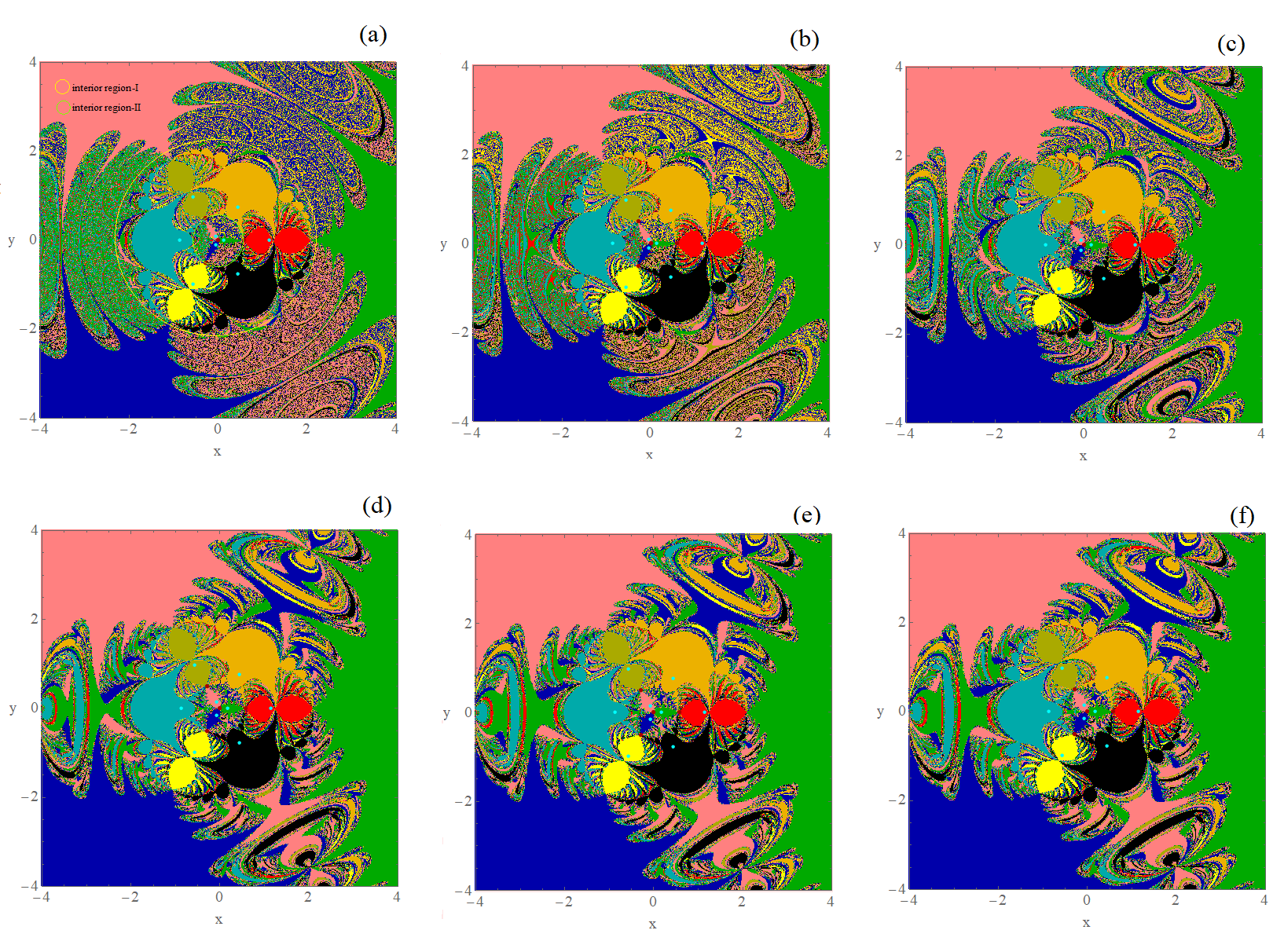}}
\caption{The Newton-Raphson basins of attraction on the configuration
$(x, y)$ plane for the increasing value of the radiation parameter $q_0$, when nine libration points exist for $\mu=0.9$, $\epsilon'=0$, $q_i=1, i=1,2,3$; (a) $q_0=0.20$; (b) $q_0=0.25$; (c) $q_0=0.5$; (d) $q_0=0.85$; (e) $q_0=0.95$; (f) $q_0=0.99$. The \emph{crimson} colour dots show the positions of the four primaries while the\emph{ teal} colour dots show the positions of the nine libration points. For the colour code denoting the attractors see the text. }
\label{Fig:B1}
\end{figure*}
The zero velocity surface (ZVS) is a  three-dimensional surface defined by the  relation $2\Omega(x, y, z) = C$. This ZVS provided the freedom to illustrate the energetically possible orbits of the fifth body whereas the Hill's regions are defined as the projections of these surfaces on the configuration $(x, y)$ plane. Moreover, the zero velocity curves (ZVCs) are the locus where the kinetic energy vanishes, describe the boundaries of the Hill's regions. Thus, we revealed that how the topology of the ZVSs, and of course the linked ZVCs change as the function of the parameters $q_0, \epsilon'$ and Jacobian constant $C$.

In Fig.\ref{Fig:Z1}, we have illustrated the zero velocity curves for varying value of the parameter $\epsilon'$ and for fixed value of  mass parameter $\mu$ where none of the primary is radiating. We have fixed the value of the mass parameter $\mu$ in the interval for which there exist fifteen libration points and also the value of the Jacobian constant $C$ is  taken constant. It is observed that when $\epsilon'=0$ (see Fig. \ref{Fig:Z1}a),  i.e., the classical case of restricted five-body problem, the primaries are surrounded by the oval shaped region, inside which the fifth particle is free to move, and consequently, the fifth body (test particle) can not communicate from one primary to either of the primary also the test particle is restricted to move from inner regions to outer region and vice-versa, i.e., there is no communication channels. In Fig. \ref{Fig:Z1}b, when $\epsilon'=0.15$, the channel at the libration points $L_{x2}, L_{xy3}$ and $ L_{xy4}$ open and consequently the test particle can orbit to communicate between the primaries $P_i, i=0,1,2,3$ inside the interior region but it cannot move to exterior region from it. Further increase in the value of the parameter $\epsilon'$ to 0.153, the three more channels around the libration points $L_{x1,5,6}$ appear so that the infinitesimal mass can communicate from interior region to the exterior region and vice-versa. In the following panel, it is observed that as the  parameter $\epsilon'$ increases to 0.52, the forbidden region shrinks significantly and constitutes three branches, each  containing $L_{x3,4,5}$,  $L_{xy1,7,9}$,  and $L_{xy2,8,10}$, respectively.  Moreover, for $\epsilon'=0.59$, three limiting situations at the libration points $L_{x5}, L_{xy9}$ and $L_{xy10}$ occur while these forbidden regions further shrink and constitute six branches each containing one  libration point for $\epsilon'=0.9$ and these forbidden regions do not disappear completely even for higher value of the parameter $\epsilon'$.  In  Fig. \ref{Fig:Z1}g and h, the regions of possible motions are illustrated for the specific value of the mass parameter $\mu$ for which nine libration points exist and observed that as the value of $\varepsilon'$ increases, the regions of possible motion also increase but even for higher value of the $\varepsilon'=0.8$, the forbidden regions do not disappear completely.

It can be concluded that as the value of the parameter $\epsilon'$, which occur due to effect of small perturbation in the centrifugal force, increases the forbidden region decreases  significantly and consequently the test particle can move freely on the entire configuration $(x, y)$ plane except the six small islands shaped forbidden region.

In Fig. \ref{Fig:Z2}, the zero velocity curves are drawn for the increasing value of the Jacobian constant. It is observed that as the value of the Jacobian constant increases, the regions of possible motion also decrease.
\section{The basins of convergence}
\label{Sec:5N}
In this section, we describe how the topology of the basins of convergence linked with the coplanar  libration points (i.e., the libation points lie on $(x, y)$ plane) are affected by the effect of small perturbations  in the Coriolis and centrifugal forces and by the radiation parameters $q_i, i=0,1,...,3$. We have adopted the same philosophy and procedure discussed in Ref. \cite{ZS17},  and  \cite{sur19}. The basin of convergence associated with the libration points are collections of the set of points which converge to specific attractor after successive iterations. We use the multivariate version of Newton-Raphson iterative method, which is defined by the map
\begin{figure*}
\centering
\resizebox{\hsize}{!}{\includegraphics{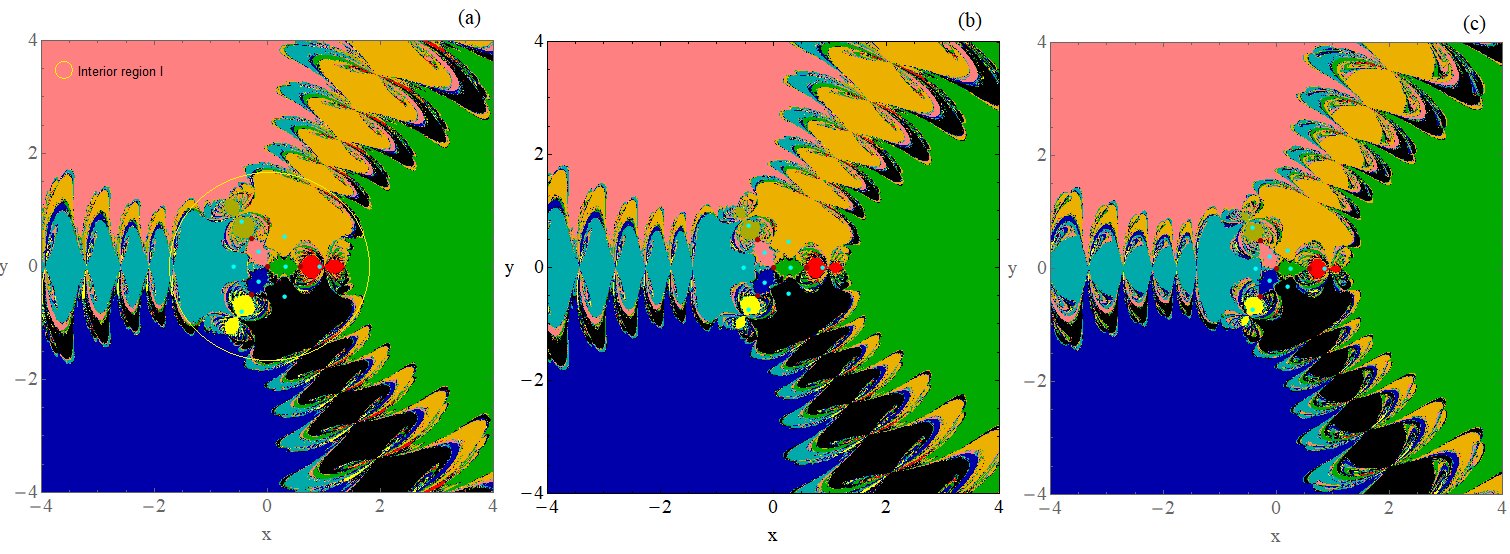}}
\caption{The Newton-Raphson basins of attraction on the configuration $(x, y)$ plane for the increasing value of the parameter $\epsilon'$, when nine libration points exist for  $\mu=0.4$; (a) $q_0=1, \epsilon'=0.2$; (b) $q_0=1, \epsilon'=0.6$; (c) $q_0=0.45, \epsilon'=0.8$. The \emph{crimson} colour dots show the positions of the four primaries while the \emph{teal} colour dots show the positions of the nine libration points. For the colour code denoting the attractors see the text.}
\label{Fig:B2}
\end{figure*}
\begin{equation}\label{Eq:B1}
 \mathbf{x}_{n+1}=\mathbf{x}_{n}-\mathbf{J}^{-1}f(x_{n}),
\end{equation}
where $\mathbf{x}=(x, y), f(\mathbf{x}_n)$ define the system of equations \ref{E:6a}-\ref{E:6b}, and $\mathbf{J}^{-1}$ is the Jacobian matrix.

In this section, we will illustrate how the radiation parameter $q_i, i=0,1,2,3$ and the parameter $\epsilon'$ which occurs due to effect of  small perturbation in the centrifugal force, affect the topology of the Newton-Raphson basins of convergence linked with the libration points (which act as  attractors) in the circular photogravitational restricted five-body problem. We have used the color-coded diagrams to classify the nodes on the configuration $(x, y)$ plane. The different colour is assigned to each pixel according to the final state of the associated initial condition.

In Fig. \ref{Fig:CP1}, the crooked path line is illustrated by the successive approximation points. The crooked path leads to a desired position of the equilibrium point when the iterative scheme converges for the particular initial condition. The collections of all initial conditions, which lead to same specific libration point, compose a basins of convergence or attracting domain.

The colour code for various attractors are as follows: $L_{x1}\rightarrow$ red, $L_{x2}\rightarrow$ cyan, $L_{x3}\rightarrow$ teal, $L_{x4}\rightarrow$ gray, $L_{x5}\rightarrow$ crimson, $L_{xy1}\rightarrow$ amber, $L_{xy2}\rightarrow$ black, $L_{xy3}\rightarrow$ pink, $L_{xy4}\rightarrow$ indigo, $L_{xy5}\rightarrow$ olive, $L_{xy6}\rightarrow$ yellow, $L_{xy7} \rightarrow$ lime green, $L_{xy8}\rightarrow$ purple, $L_{xy9}\rightarrow$ light blue, $L_{xy10}\rightarrow$ orange, and the non-converging points are shown in white colour.

In Fig. \ref{Fig:B1}, we have depicted the basins of convergence associated with the co-planar libration points by using the multivariate version of the Newton-Raphson iterative scheme for six increasing values of the radiation parameter $q_0$ i.e., when the central primary is radiating and also $q_i=1, i=1,2,3$ i.e., the other three primaries are non-radiating bodies.  In an attempt to observe the effect of the radiation parameter $q_0$, the mass parameter $\mu$ is fixed in the interval for which there is only nine libration points.  It is observed that in all the panels the extent of the basins of convergence associated with each of  the libration point is infinite however the configuration $(x, y)$ plane is fully covered by well-formed domain of the basins of convergence. Additionally, the vicinity of the basins boundaries are composed of the mixture of the initial conditions which have fractal like geometry. It is observed that the slight change in the radiation parameter $q_0$  leads to the significant change in the structure of the basins of convergence. In Fig. \ref{Fig:B1}a, the basins of convergence is illustrated for parameter $q_0=0.2$, i.e., the central primary is source of radiation and the area  enclosed by schematic yellow circle will be named as interior region-I, and area  enclosed by schematic green circle will be named as interior region-II, from now. It is observed that, outside the interior region-I,  three wings shaped area is very noisy. The domain of the basins of convergence associated with the libration points  $L_{x1}$, and $L_{xy5,6}$  look like three exotic bugs, with many legs and antenna whereas the domain of the basins of convergence $L_{x3}$ and $L_{xy1,2}$ look like butterfly wings, which lie inside the region-I,  are regular but their boundaries are separated by  thin strips which are composed of initial conditions. The region-II is also noisy except the three leaves shaped domain of the basins of convergence associated with the libration points $L_{x2}$ and $L_{xy3, 4}$.
\begin{figure*}
\centering
\resizebox{\hsize}{!}{\includegraphics{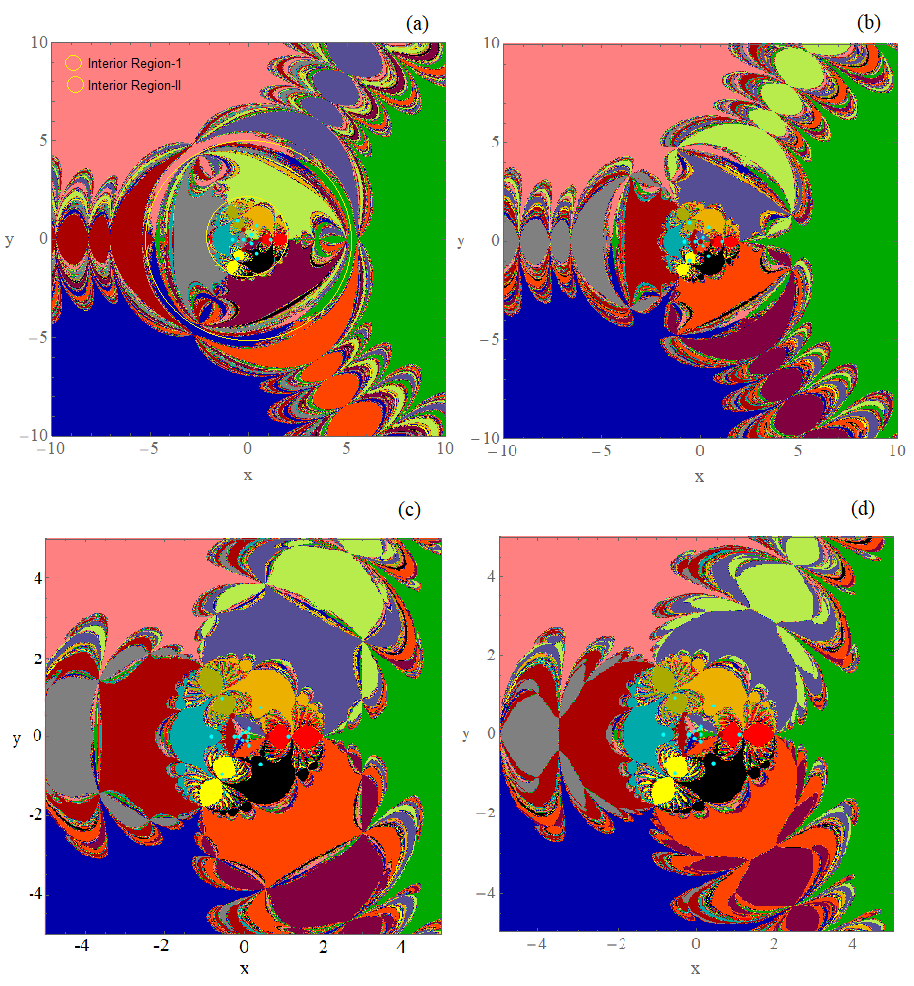}}
\caption{The Newton-Raphson basins of attraction on the configuration
$(x, y)$ plane for the increasing value of the radiation parameter $q_0$, when fifteen libration points exist for $\mu=0.98124858$; $\epsilon=0.25$; $q_i=1,  i=1, 2, 3$; (a) $q_0=0.09$; (b) $q_0=0.15$; (c) $q_0=0.55$; (d) $q_0=0.95$. The \emph{crimson} colour dots show the positions of the four primaries while the \emph{ teal} colour dots show the positions of the fifteen libration points. For the colour code denoting the attractors see the text.}
\label{Fig:B3}
\end{figure*}
\begin{figure*}
\centering
\resizebox{\hsize}{!}{\includegraphics{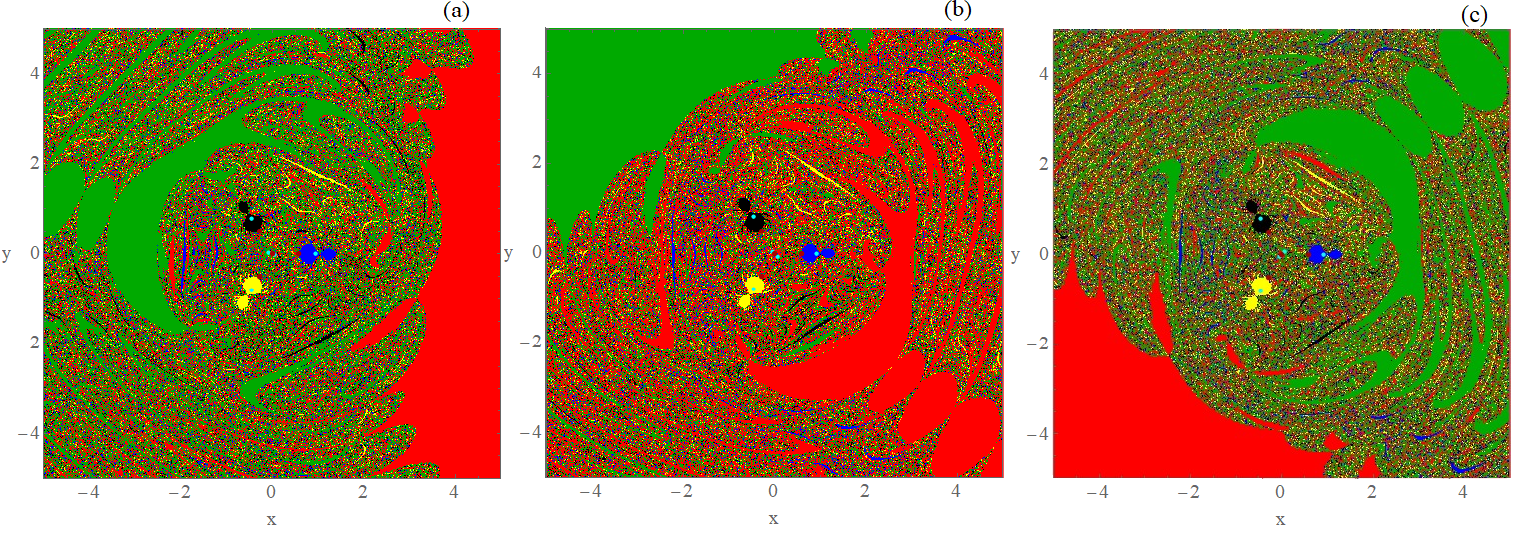}}
\caption{The Newton-Raphson basins of attraction on the configuration
$(x, y)$ plane  when five libration points exist for $\mu=0.98124858$;  $\epsilon'=0.25$; (a) $q_1=0.45;$ $q_2=0.35$; $q_3=0.4$;  $q_0=0.15$; (b) $q_1=0.35$; $q_2=0.45$; $q_3=0.4$; $q_0=0.15$; (c) $q_1=0.35$; $q_2=0.4$; $q_3=0.45$; $q_0=0.15$. The \emph{crimson} colour dots show the positions of the four primaries while the\emph{ teal} colour dots show the positions of the five libration points. For the colour code denoting the attractors is as follows: $L_{xy1}$ (\emph{blue}); $L_{xy2}$ (\emph{green}); $L_{xy3}$ (\emph{red}); $L_{xy4}$ (\emph{black}); $L_{xy5}$ (\emph{yellow}); the non-converging points (white).}
\label{Fig:B4}
\end{figure*}
Moreover, as the value of radiation parameter $q_0$ increases, the three wings shaped regions, which were noisy in previous panel and looks like chaotic sea of initial conditions, now started to convert into some small regular islands.  In Fig. \ref{Fig:B1}c-f, we can observe that a large area of these wings shaped regions converted to well shaped basins of convergence as the value of radiation parameter $q_0$ increases. Additionally, the three leaves  shaped domain of the  basins of convergence inside the region-II increase and most of the region inside it, is now covered by well formed basins of convergence.

In Fig. \ref{Fig:B2}, the basins of convergence are illustrated as the function of parameter $\epsilon'$ when nine libration points exist. It can be observed that the entire configuration plane is covered by well formed basins of attraction, however the domain of each of the basins of convergence has infinite extent.  As we increase $\epsilon'$, the interior region-I, shrinks whereas the number of leaves in the wings shaped region increases. Moreover, we observe the vicinity of the basins boundaries are composed of the mixture of initial conditions. This fact unveils that the basins boundaries are highly chaotic and hence, the initial conditions, inside these areas are extremely sensitive to its final state. Indeed, even a very slight change in the value of initial conditions will lead to a entirely unexpected attractor.  The domain of the basins of convergence associated with the libration points $L_{x1}, L_{xy5, 6}$ look like exotic bugs with many legs and antenna, shrink as the parameter $\epsilon'$ increases.

In Fig. \ref{Fig:B3}, the basins of convergence are depicted for varying values of the radiation parameter $q_0$, when fifteen libration points exist. The configuration plane is covered by well-formed basins of attraction whereas the basins boundaries are composed of the mixtures of the initial conditions.  When we compare the Figs. \ref{Fig:B3}a and b, it can be observed that the three heart shaped regions, which occur at the joining point of wing shaped area, separate into two parts and consequently hinges which connect the interior region-I and the wings shaped region shrinks. Moreover, the area inside the interior region-II becomes more regular in comparison of panel-b. Furthermore, the increase in the parameter $q_0=0.95$ the basins boundaries which connect the leaves of the wings are now regular and the chaotic basin boundaries disappear.

The basins of the convergence associated with the libration points, for the case when all the primaries are radiating and the values of these radiation parameters are different,  are illustrated in Fig. \ref{Fig:B4}. As we have observed that in this case the symmetry of the libration points about the $x-$axis is destroyed and only  five non-collinear libration points exist, the basins of convergence  also look strange. We observe that a majority of the $(x , y)$ plane is covered by the highly fractal\footnote{The term fractal simply unveils the fact that the specific area has a fractal-like geometry, without conducting extra calculations, as in \cite{AGU01}.} mixture of initial conditions. The only regular island shown in\emph{ green} and \emph{red} colour corresponds to those libration points which exist near the origin, i.e., $L_{xy2,3}$ (see  Fig. \ref{Fig:S1}).  However, the basins of convergence associated with the libration points $L_{xy1, 4,5}$ look like exotic bugs without legs and antennas.  The  Fig. \ref{Fig:B4}b and c    look like mirror images about the $x-$axis when the green and red colors are interchanged. This happens only  because the value of the radiation parameter $q_2$ and $q_3$ are interchanged.
\section{Discussions and conclusions}
\label{Sec:6}
In the present manuscript, we have successfully illustrated the influence of the effect of small perturbations in the Coriolis and centrifugal forces on the positions of the libration points, their stability and the regions of possible motion  in context of  photogravitational version of restricted five-body problem.
We have further, investigated numerically  the basins of convergence, linked to the libration points, in the problem.  We have used the multivariate version of the Newton-Raphson iterative scheme  to unveil the associated basins of attraction on the configuration $(x, y)$ plane. The role of the attracting  domains is very crucial as it can explain how each initial point of the $(x, y)$ plane is attracted by the attractors of the dynamical system, which are, indeed, a libration point of the system. We have  successfully managed to detect how the geometry of the Newton-Raphson basins of attraction changes as a function of the radiation parameters and the $\epsilon'$.

 In the problem, we have taken either the central or all of the primaries as source of radiations. The main conclusions can be summarized as follows:
\begin{itemize}
  \item The number as well as the positions of the libration points are highly sensitive with the change in the value of the parameter $\epsilon'$. The  value of $\mu^*$, ( i.e., the value of mass parameter is critical value since it delimits the point where the number of the equilibrium points changes) decreases as  $\epsilon'$ increases.
      \item  If only the central primary radiates,  i.e., $q_i=1, i= 1, 2, 3,$ and other perturbations are neglected, all the libration points move away along a straight line from the central primary when $q_0\in(0, 1]$ whereas, all the libration points move towards it along a straight line if the effect in the centrifugal force increases where none of the primary is radiating.
      \item The non collinear libration points $L_{xyi}, i=1,2,3,4$ move away from the  central primary along a curve whereas the other libration points move along the straight line when the radiation parameter $q_0$ increases where the other primaries are also the source of radiation such that $q_1\neq q_2=q_3$.
      \item The stability analysis of the libration points have unveiled that in most of the cases only the collinear libration point $L_{x3}$ is stable. Moreover, for various combination of parameters, the libration points $L_{x3,4,5}$ and $L_{xy7,8,9,10}$ are also stable sometime or another. It is interesting to note that only those non-collinear libration points are stable, which lie in the first or fourth quadrant.
      \item If we compare our result from the classical five-body problem (see \cite{oll88}), it is observed that the critical value of $\beta$ decreases significantly when the effect of the Coriolis and centrifugal forces is taken into consideration.
      \item If we increase the parameter $\epsilon'$, the possible regions of motion increase whereas they decrease when the Jacobian constant $C$ increases. Further, these regions are unaffected by the change in value of $\epsilon$, i.e., parameter occur due to presence of the Coriolis force as the effective potential is independent from this parameter.
      \item The basins of convergence associated with all the libration points have infinite extent. The the interior region-I,  interior region-II, and the wing shaped region become more regular as the radiation parameter due to the central primary increases. Moreover, the entire basins of convergence shrinks in whole as the parameter $\epsilon'$ increases while the basins of convergence remains unaffected with the change in the value of $\epsilon$ which control the Coriolis force.
      \item The majority of the area of basins of convergence look like chaotic sea, composed of  different type of initial conditions when the value of the radiation parameters $q_i, i=0,1,..,3$ are different. Therefore, in these cases it is almost impossible to predict that which initial conditions will converge to which of the libration points.
\end{itemize}
The latest version \emph{11.3} of Mathematica$^\circledR$ is used for the graphical illustration of the paper. It is  interesting to use other types of iterative schemes in the present problem to discuss the similarities and differences in the context of the basins of convergence linked with the libration points of the dynamical system. We believe that  the above-mentioned results and ideas will be useful in the active field of attracting domains of libration points in various dynamical system.

\section*{Compliance with Ethical Standards}
\begin{description}
  \item[-] Funding: The authors state that they have not received any research
grants.
  \item[-] Conflict of interest: The authors declare that they have no conflict of
interest.
\end{description}
\bibliographystyle{aps-nameyear}

\begin{thebibliography}{}

 \bibitem{AAG16} Abouelmagd E. I., Alzahrani F., Guiro J. L. G., Hobiny A., Periodic orbits around the collinear libration points. \emph{J. Nonlinear Sci. Appl. (JNSA).} (2016) 9 (4): 1716 -1727.
 \bibitem{AG16} Abouelmagd E. I., Guirao J. L. G., On the perturbed restricted three-body problem.  \emph{Applied Mathematics and Nonlinear Sciences} (2016) 1 (1): 123 – 144.
 \bibitem{A17}   Ansari A. A.,  Investigation of the effect of albedo and oblateness on the circular restricted four variable bodies problem. \emph{Applied Mathematics and Nonlinear Sciences} (2017) 2(2) 529–542.
 \bibitem{AG19}  Abouelmagd E. I., Guirao J. L.G., Llibre J. Periodic orbits for the perturbed planar circular restricted 3-body problem. \emph{Discrete and Continuous Dynamical Systems - Series B (DCDS-B)} (2019) 24 (3) 1007-1020.
\bibitem{agg18}
Aggarwal, R.,  Mittal, A.,   Suraj, M. S.,   Bisht, V., The effect of small perturbations in the Coriolis and centrifugal forces on the existence of libration points in the restricted four-body problem with variable mass. \emph{Astronomical notes}, \textbf{339}(6)  (2018) 492-512. doi.org/10.1002/asna.201813411.
\bibitem{AGU01}Aguirre, J., Vallejo, J.C., Sanju\'{a}n, M.A.F., Wada basins and chaotic invariant sets in the H\'{a}non-Heiles system. \emph{Phys. Rev.} E 64, 066208 (2001)
\bibitem{AAGH17}Alzahrani F., Abouelmagd E. I., Guirao J. L.G., Hobiny A.  On the libration collinear points in the restricted three-body problem. \emph{Open Physics} (2017) 15 (1): 58-67.
\bibitem{bha78}
    Bhatnagar, K.B., Hallan, P.P., Effect of perturbations in Coriolis and centrifugal forces on the stability of libration points in the restricted problem. \emph{Celestial Mechanics}, \textbf{18} (1978)  105-112. DOI: 10.1007/BF01228710
\bibitem{bha83}
    Bhatnagar, K.B., Hallan, P.P., The effect of perturbations in Coriolis and centrifugal forces on the nonlinear stability of equilibrium points in the restricted problem of three bodies. \emph{Celestial Mechanics}, \textbf{30} (1983) 97. doi:10.1007/BF01231105
\bibitem{EAKP16}Elshaboury S. M., Abouelmagd E. I., Kalantonis V.S., Perdios  E. A. The planar restricted three-body problem when both primaries are triaxial rigid bodies: Equilibrium points and periodic orbits. \emph{Astrophys. Space Sci.}, \textbf{361(9)} (2016)315
\bibitem{gao17} Gao, C., Yuan, J., Sun, C., Equilibrium points and zero velocity surfaces in the axisymmetric restricted five-body problem, \emph{Astrophys. Space Sci.}, \textbf{362} (2017) 72
 \bibitem{K99}Kalvouridis, T.J.,  A planar case of the $n+1$ body problem: the "ring"  problem.   \emph{Astrophys. Space Sci.} \textbf{260}, 309-325 (1999)
  \bibitem{M90}Maxwell, J.C.: On the Stability of the Motion of Saturn’s Rings. Scientific Papers of James Clerk Maxwell,
Vol. 1, p. 228. Cambridge University Press, Cambridge (1890)
\bibitem{pp13} Papadouris, J.P., Papadakis, K.E., Equilibrium points in the photogravitational restricted four-body problem. \emph{Astrophys. Space Sci.}, \textbf{344} (2013)  21-38
\bibitem{p16} Papadakis, K.E., Families of three dimensional periodic solutions in the circular restricted four-body problem. \emph{Astrophys. Space Sci.}, \textbf{361} (2016) 129
  \bibitem{PK07} Papadakis, K.E., Kanavos, S.S., Numerical      exploration of the photogravitational restricted five-body problem. \emph{Astrophys. Space Sci.}, \textbf{310}  (2007) 119--130
\bibitem{PE17} Pathak N., Elshaboury S. M., On the triangular points within frame of the restricted three–body problem when both primaries are triaxial rigid bodies. \emph{Applied Mathematics and Nonlinear Sciences} (2017) 2(2) 495–508.
\bibitem{oll88} Oll\"{o}ngren, A., On a particular restricted five-body problem, an analysis with computer algebra.\emph{ J. Symb. Comput.}, \textbf{6 }  (1988) 117--126
\bibitem{RZ99} Ragos, O., Zagouras, C.,  Periodic solutions around the collinear Lagrangian points in the photogravitational restricted three-body problem: Sun-Jupiter case.   \emph{Celest. Mech. Dynam. Astron.} \textbf{50}, 325-347 (1999)
\bibitem{SGA19}Selim H. H., Guirao J. L.G., Abouelmagd E I.  Libration points in the restricted three-body problem: Euler angles, existence and stability. \emph{Discrete and Continuous Dynamical Systems - Series S (DCDS-S) }(2019) 12 (45): 703 – 710.
\bibitem{sin15} Singh, J., Vincent, A.E., Effect of perturbations in the Coriolis and centrifugal forces on the stability of equilibrium points in the restricted four-body problem, \emph{Few-Body Syst.}, \textbf{56} (2015) 713--723. DOI 10.1007/s00601-015-1019-3
\bibitem{sur19} Suraj, M.S., Sachan, P., Zotos, E.E., Mittal. A., Aggarwal, R., On the fractal basins of convergence of the libration points in the axisymmetric five-body problem: The convex configuration, \emph{Int. J.  of Non-Linear Mechanics,} \textbf{109} (2019a) 80-106
\bibitem{sur19b} Suraj, M.S., Sachan, P., Zotos, E.E., Mittal. A., Aggarwal, R., On the Newton–Raphson basins of convergence associated with the libration points in the axisymmetric restricted five-body problem: The concave configuration, \emph{Int. J.  of Non-Linear Mechanics,} \textbf{112} (2019b) 25-47
\bibitem{Sur19c} Suraj, M.S.,    Abouelmagd, E. I., Aggarwal, R.,  Mittal, A., The analysis of restricted five--body problem within frame of variable mass,  \emph{New Astronomy,} 70, 12-21(2019c) https://doi.org/10.1016/j.newast.2019.01.002
\bibitem{Sur19d} Suraj, M.S.,    Sachan, P.,   Mittal, A., Aggarwal, R., The effect of small perturbations in the Coriolis and centrifugal forces in the axisymmetric restricted five-body problem,
\emph{Astrophys Space Sci} (2019d) 364: 44.
\bibitem{Sur17} Suraj, M.S.,    Aggarwal, R.,   Arora, M., On the restricted four-body problem  with the effect of small perturbations in the Coriolis and centrifugal forces,  \emph{Astrophys. Space Sci.} \textbf{362} (2017a) 159
\bibitem{Sur17b} Suraj, M.S., Asique, M.C., Prasad, U. et al., Fractal basins of attraction in the restricted four-body problem when the primaries are triaxial rigid bodies.  \emph{Astrophys Space Sci.,}\textbf{362}, 211 (2017b).
\bibitem{SMA18}Suraj, M.S., Zotos, E.E.,  Aggarwal, R., Mittal, A.,
Unveiling the basins of convergence in the pseudo-Newtonian planar circular restricted four-body problem, \emph{New Astronomy,} \textbf{66}, 52--67,  (2018a)
\bibitem{SZK18} Suraj, M.S., Zotos, E.E., Kaur, C., Aggarwal, R., et al., Fractal basins of convergence of libration points in the planar Copenhagen problem with a repulsive quasi-homogeneous Manev--type potential.  \emph{Int. J. Non-Linear Mech.}, \textbf{103}, 113-127, (2018b)
%
\bibitem{SAA18} Suraj, M.S., Mittal, A., Arora, M. et al., Exploring the fractal basins of convergence in the restricted four-body problem with oblateness.  \emph{Int. J. Non-Linear Mech.}, \textbf{102}, 62--71 (2018c)
\bibitem{sze67}Szebehley, V., Stability of the points of equilibrium in the restricted problem. \emph{Astron. J.}, \textbf{72} (1967) 7
\bibitem{ZS17} Zotos, E.E., Suraj, M.S.,  Basins of attraction of equilibrium points in the planar circular restricted five-body problem, \emph{Astrophys. Space. Sci.,} \textbf{363}  (2017) 20
\bibitem{Z18} Zotos, E.E., On the Newton--Raphson basins of convergence of the out-of-plane equilibrium points in the Copenhagen problem with oblate primaries. \emph{Int. J. Non-Linear Mech.}, \textbf{103}, 93--105 (2018)
\bibitem{ZSMA18} Zotos, E.E., Suraj, M.S., Jain, M., Aggarwal, R., Revealing the Newton-Raphson basins of convergence in the circular pseudo-Newtonian Sitnikov problem.  \emph{Int. J. Non-Linear Mech.}, \textbf{105}, 43--54 (2018).
\end{thebibliography}

\end{document}